\documentstyle[aps]{revtex}
\begin{document}
\draft
\author{B.V.Ivanov}
\title{Towards a regular type N vacuum gravitational field}
\address{Institute for Nuclear Research and Nuclear Energy\\
Tzarigradsko Shausse 72, Sofia 1784, Bulgaria}
\date{5 June 1999}
\maketitle

\begin{abstract}
An exact twisting type N vacuum solution is found. It has gauge and
curvature invariants which are regular in the angular coordinates and decays
to flat spacetime for big retarded times.
\end{abstract}


The study of gravitational radiation far away from bounded sources becomes
more and more important. It is governed by the Petrov type N part of the
Riemann tensor and exact expanding vacuum solutions of type N seem to play
fundamental role in the understanding of this phenomenon. All non-twisting
solutions are known [1,2] but have singularities which extend to spatial
infinity (singular pipes). This is a typical feature of the class of
Robinson-Trautman solutions [3]. For the twisting case only the Hauser
solution is known [4,5] but it is singular too [6]. A linearised twisting
solution also possesses singular pipes [7], but a third and fourth order
iterations of it turn out to be regular [8,9]. However, higher orders become
singular again [9] and even the third order iteration leads to singularities
in the recently discovered curvature invariant [10,11].

In this work we present an exact twisting solution which satisfies the
existing criteria for regularity. We shall work in the tetrad and coordinate
system used in [1]. The metric is 
\begin{equation}
ds^2=\frac{2d\zeta d\bar \zeta }{\rho \bar \rho P^2}-2\Omega \left(
dr+Wd\zeta +\bar Wd\bar \zeta +N\Omega \right)  \label{1}
\end{equation}
\[
\Omega =du+Ld\zeta +\bar Ld\bar \zeta 
\]
where $r$ is the coordinate along a null congruence, $u$ is the retarded
time and $\zeta $ is connected to the usual spherical angles by 
\begin{equation}
\zeta =\sqrt{2}\tan \frac \theta 2e^{i\varphi }  \label{2}
\end{equation}
The metric components are determined by a real function $P$ and a complex
function $L$ which are $r$-independent: 
\begin{equation}
\rho =-\left( r+i\Sigma \right) ^{-1}  \label{3}
\end{equation}
\begin{equation}
W=\rho ^{-1}L_u+i\partial \Sigma  \label{4}
\end{equation}
\begin{equation}
N=-r\left( \ln P\right) _u+K/2  \label{5}
\end{equation}
where $\partial =\partial _\zeta -L\partial _u$ and the gauge invariants $%
\Sigma /r$ and $K/r^2$ will be written out later, after their
simplification. The basic functions $P$ and $L$ satisfy the field equation 
\begin{equation}
Im\partial \partial \bar \partial \bar \partial V=0  \label{6}
\end{equation}
where $V_u=P$ and the condition that the third Weyl scalar $\Psi _3$
vanishes and hence the solution is of type N or higher: 
\begin{equation}
\partial \psi ^{-1}=0  \label{7}
\end{equation}
\begin{equation}
\psi ^{-1}=P^{-1}\left( \bar \partial \bar \partial V\right) _u  \label{8}
\end{equation}

We shall use an adaptation of Stephani's method [12] which was proposed
originally for type II solutions. Let us choose as a basic element the
complex potential $\psi \left( \zeta ,\bar \zeta ,u\right) $ which satisfies 
$\partial \psi =0$. This immediately gives 
\begin{equation}
L=\psi _\zeta /\psi _u  \label{9}
\end{equation}
and $\psi $ may be taken as an arbitrary function $\psi =\psi \left( \zeta ,%
\bar \zeta ,u\right) $. This relation can be inverted to obtain $u=u\left(
\zeta ,\bar \zeta ,\psi \right) $. This allows to consider $V$ as a function
of $\zeta ,\bar \zeta $ and $\psi $, $V=V\left( \zeta ,\bar \zeta ,u\left(
\zeta ,\bar \zeta ,\psi \right) \right) =V\left( \zeta ,\bar \zeta ,\psi
\right) $. Then $\partial V=V_\zeta $ and $\left[ \partial ,\partial _\psi
\right] =0$. Eqs. (7) and (8) become 
\begin{equation}
\bar \psi H_{\zeta \zeta }=H  \label{10}
\end{equation}
\begin{equation}
P=i\psi _uH  \label{11}
\end{equation}
where $iH=V_\psi $. In order to simplify Eqs. (6) and (10) we discuss the
subclass of functions $\psi $ with $u$-independent real part: 
\begin{equation}
\psi =q\left( \zeta ,\bar \zeta \right) -ih\left( \zeta ,\bar \zeta
,u\right)   \label{12}
\end{equation}
\[
\bar \psi =2q\left( \zeta ,\bar \zeta \right) -\psi 
\]
Then Eq. (11) shows that $H$ is real and replacing $\bar \psi $ from Eq.
(12) into Eq. (10) we can determine $H=H\left( \zeta ,\bar \zeta ,\psi
\right) $. Eqs. (9) and (11) give $P$ and $L$ . A non-trivial $P$ demands $%
H\neq 0,$ $h_u\neq 0$. Eq. (12) also gives $\bar \partial V=V_{\bar \zeta
}+2q_{\bar \zeta }V_\psi $. Applying the commutator $\left[ \partial ,\bar 
\partial \right] =2Q\partial _\psi $, where $Q=q_{\zeta \bar \zeta }$, to
Eq. (6) one obtains 
\begin{equation}
2Q\left( H_{\zeta \bar \zeta }+QH_\psi +2q_{\bar \zeta }H_{\psi \zeta
}\right) +Q_{\bar \zeta }H_\zeta +Q_\zeta H_{\bar \zeta }+2q_{\bar \zeta
}Q_\zeta H_\psi +\frac 12Q_{\zeta \bar \zeta }H=0  \label{13}
\end{equation}

The twist $\Sigma $ and the other gauge invariant $K$ (since we always take $%
r$ as directly given and independent of $\zeta $ [7,8]) read 
\begin{equation}
\Sigma =h_uH^2Q  \label{14}
\end{equation}
\begin{equation}
K=h_u^2H\left( \partial \bar \partial +\bar \partial \partial \right) \ln H
\label{15}
\end{equation}
We must ensure that $Q\neq 0$ for a non-zero twist. The only non-trivial
Weyl scalar is 
\begin{equation}
\Psi _4=i\rho h_u^3H^2\psi ^{-2}  \label{16}
\end{equation}
The non-vanishing of $h_u$ and $H$ guarantees that $\Psi _4\neq 0$ and the
solution is of type N. The true curvature invariant $I$ of Bi\v c\'ak and
Pravda [10] reads 
\begin{equation}
\sqrt{I}=48\left( \rho \bar \rho \right) ^2\Psi _4\bar \Psi _4  \label{17}
\end{equation}

Instead of Eqs. (6) and (7) we have derived a system of two linear second
order with respect to $H$ equations (10) and (13). The procedure of finding
a solution is the following. We fix $q$ and take an arbitrary $h$. $L$ is
given by Eq. (9). We substitute from Eq. (12) $\bar \psi \left( \zeta ,\bar 
\zeta ,\psi \right) $ in Eq. (10) and solve it for $H,$ $\psi $ being
treated as a parameter. Then we try to satisfy Eq. (13). $P$ is found from
Eq. (11). A possible problem is the reality of $H$. The system (10) and (13)
is simpler than the numerous non-linear high order equations found for $L$
in the gauge $P=1$ when Killing or homothetic Killing vectors are present
[13-16].

Let us find next a solution $H\left( x,u\right) $, where $x=\zeta +\bar \zeta
$, possessing the simplest possible twist $Q=Q_0=const,$ $q=Q_0x^2/2$. Eq.
(10) becomes 
\begin{equation}
\left( Q_0x^2-\psi \right) H_{xx}=H  \label{18}
\end{equation}
Introduction of the variable $z=Q_0x^2/\psi $ transforms Eq. (18) into 
\begin{equation}
z\left( 1-z\right) H_{zz}+\frac 12\left( 1-z\right) H_z+H/4Q_0=0  \label{19}
\end{equation}
This is a hypergeometric equation and one of its fundamental solutions is $%
F\left( a,b,c,z\right) $ where $a\left( 1+2a\right) =1/2Q_0,$ $b=-a-1/2,$ $%
c=1/2$. We must make $H$ real, having in mind that $\bar z=z/\left(
z-1\right) $ and $\bar \psi =\psi \left( z-1\right) $. The necessary linear
transformations of $F$ exist when $c=2b$, which means $Q_0=4/3$ [17]. Then 
\begin{equation}
H_1=\left( -i\psi \right) ^{3/4}F\left( -\frac 34,\frac 14,\frac 12,z\right)
\label{20}
\end{equation}
\begin{equation}
z=4x^2/3\psi  \label{21}
\end{equation}
\begin{equation}
\psi =\frac 23x^2-ih  \label{22}
\end{equation}
is one real solution of Eq. (19). The expressions $c-a$ and $c-b$ are not
integers and the second real solution is 
\begin{equation}
H_2=x\left( i\psi \right) ^{1/4}F\left( -\frac 14,\frac 34,\frac 32,z\right)
\label{23}
\end{equation}
Both solutions may be expressed also by Legendre functions because $c=2b$
[17]. They satisfy an identity which follows from Eq. (21): 
\begin{equation}
\frac 34H=\psi H_\psi +\frac 12xH_x  \label{24}
\end{equation}
When $Q=Q_0$ and $\partial _\zeta =\partial _x$ Eq. (13) becomes 
\begin{equation}
H_{xx}+Q_0H_\psi +2Q_0xH_{\psi x}=0  \label{25}
\end{equation}
Exploiting Eqs. (18) and (24) we see that $H_{1,2}$ satisfy Eq. (25) for $%
Q_0=4/3$. In total, we have found two type N solutions depending on an
arbitrary real function $h\left( x,u\right) $ and given by 
\begin{equation}
P_{1,2}=h_uH_{1,2}  \label{26}
\end{equation}
\begin{equation}
L=4ix/3h_u+h_x/h_u  \label{27}
\end{equation}

The function $h$ may be chosen in such a way that $\Sigma ,$ $K,$ $\Psi _4$
and $I$ remain regular in $x$ for $-\infty \leq x\leq \infty $, i.e. possess
no singular pipes. Namely, let us take 
\begin{equation}
h=g\left( u\right) \left( 1+x^4\right) ^{-1}  \label{28}
\end{equation}
where $g\neq 0$ and is bounded, but otherwise arbitrary function. Then 
\begin{equation}
\Sigma =4H^2g_u/3\left( 1+x^4\right)  \label{29}
\end{equation}
\begin{equation}
K=-2H_x\left( H_x+8xH_\psi /3\right) \left( 1+x^4\right) ^{-8}g_u^2
\label{30}
\end{equation}
\begin{equation}
\sqrt{I}=48H^4g_u^6\left( r^2+\Sigma ^2\right) ^{-3}\left( 1+x^4\right)
^{-2}\left[ 4x^4\left( 1+x^4\right) ^2/9+g^2\right] ^{-2}  \label{31}
\end{equation}
Eqs. (21) and (22) show that $z$ is always complex and $\mid z\mid \leq 2$.
Hence, $F$ in Eqs. (20) and (23) is regular in $x$. Singular pipes do not
arise in Eqs. (29),\ (30) and (31) even when $x\rightarrow \infty $ because $%
H\sim x^{3/2},$ $H_x\sim x^{1/2}$ and $H_\psi \sim x^{-1/2}$. The condition $%
g\neq 0$ prevents the appearance of singularity in $I$ at $x=0$. Thus the
gauge invariants $\Sigma $ and $K$ and the true coordinate invariant $I$ are
regular for any $x$ i.e. on the two-dimensional sphere. The function $%
g\left( u\right) $ is characteristic for gravitational radiation since it
may bring ''news''. If $g\left( u\right) =1+k\left( u\right) $ and $%
k_u\rightarrow 0$ when $u\rightarrow \infty $, the three invariants vanish
for large retarded times.

Finally, let us make connection with previous results about twisting type N
fields. The even Hauser solution [5] separates variables in the present
coordinates [1]: 
\begin{equation}
L_H=2\left( u+i\right) /x  \label{32}
\end{equation}
\begin{equation}
P_H=x^{7/2}f\left( u\right)  \label{33}
\end{equation}
\begin{equation}
f\left( u\right) =F\left( -\frac 18,-\frac 38,\frac 12,-u^2\right)
\label{34}
\end{equation}
Eq. (32) leads to the potential 
\begin{equation}
\psi _H=Q_0x^2\left( 1-iu\right) /2  \label{35}
\end{equation}
It falls within the subclass we are discussing. A combination of linear and
quadratic transformations applied to $f\left( u\right) $ yields 
\begin{equation}
P_H=AP_1+BP_2  \label{36}
\end{equation}
where $A$ and $B$ are constants and $h=2ux^2/3$. The result for the odd
solution is similar. This is not surprising because a linear combination of
solutions of the linear equations (10) and (25) is also a solution if $\psi $
is the same. Eq. (36) fixes $Q_0=4/3$ in Eq. (35).

Formulae (29-31) are replaced by 
\begin{equation}
\Sigma _H=2x^5f  \label{37}
\end{equation}
\begin{equation}
K_H=-\frac 89x^5S\left( u\right)   \label{38}
\end{equation}
\begin{equation}
\sqrt{I_H}=108f^4x^{10}\left( 1+u^2\right) ^{-2}\left( r^2+4f^4x^{10}\right)
^{-3}  \label{39}
\end{equation}
where $S\left( u\right) $ $=\left( 9f/4-3uf_u\right) ^2+9f_u^2$. It is
clearly seen that $\Sigma _H$ and $K_H$ have singular pipes at $x=\infty $.
Quite interestingly, $I_H$ is regular in $x$. When $u\rightarrow \infty ,$ $%
f\sim u^{3/4}$ and $\Sigma _H$ and $K_H$ diverge, but $I_H\rightarrow 0$.

There are other solutions of Eq. (10) with $L$ given by Eq. (32) and $\psi $
given by Eq. (35). They correspond to rational degenerations of the
hypergeometric function for certain values of $Q_0$. When $Q_0=1/2$ and $%
Q_0=1/6$ Eq. (19) yields the solutions 
\begin{equation}
P_{C1}=-2^{-6}x^6\left( u^2+1\right)  \label{40}
\end{equation}
\begin{equation}
P_{C2}=-12^{-7/2}x^7\left( u^2+1\right)  \label{41}
\end{equation}
respectively. The constants in the above equations are irrelevant since $P$
is determined up to a multiplicative constant. These are the Collinson's
semi-solutions [18], because they do not satisfy the constraint equation
(25).

This work was supported by the Bulgarian National Fund for Scientific
Research under contract F-632.

\end{document}